\documentclass{Interspeech2024}
\usepackage{multirow}
\usepackage{subfig}

\interspeechcameraready

\title{Benchmarking Children's ASR with Supervised and Self-supervised Speech Foundation Models}

\name[affiliation={1}]{Ruchao}{Fan}
\name[affiliation={1}]{Natarajan Balaji}{Shankar}
\name[affiliation={1}]{Abeer}{Alwan}

\address{
  $^1$Dept. of Electrical and Computer Engineering, University of California, Los Angeles, U.S.A \thanks{This work was supported in part by the NSF.}}
\email{{fanruchao,balaji1312}@g.ucla.edu, alwan@ee.ucla.edu}


\begin{document}

\maketitle
\begin{abstract}
    
Speech foundation models (SFMs) have achieved state-of-the-art results for various speech tasks in supervised (e.g. Whisper) or self-supervised systems (e.g. WavLM). However, the performance of SFMs for child ASR has not been systematically studied. In addition, there is no benchmark for child ASR with standard evaluations, making the comparisons of novel ideas difficult. In this paper, we initiate and present a comprehensive benchmark on several child speech databases based on various SFMs (Whisper, Wav2vec2.0, HuBERT, and WavLM). Moreover, we investigate finetuning strategies by comparing various data augmentation and parameter-efficient finetuning (PEFT) methods. We observe that the behaviors of these methods are different when the model size increases. For example, PEFT matches the performance of full finetuning for large models but worse for small models. To stabilize finetuning using augmented data, we propose a perturbation invariant finetuning (PIF) loss as a regularization.\footnote{Our code is available at \url{https://github.com/Diamondfan/SPAPL_KidsASR}}.
\end{abstract}

\section{Introduction}

Large foundation models have increasingly gained attention in the research community because of their impressive zero-shot and in-context learning ability~\cite{wei2022emergent,li2023prompting,peng2023reproducing}. Specifically in the speech area, the Whisper-large model~\cite{Rad23whisper} has shown great robustness to diverse domains of speech data by learning from large-scale supervised data in a multi-task training setting. In addition to Whisper, another type of speech foundation models (SFM) is obtained through self-supervised learning, e.g. Wav2vec2.0~\cite{baevski2020wav2vec}, HuBERT~\cite{hsu2021hubert}, WavLM~\cite{chen2022wavlm}, and W2VBERT2.0~\cite{chung2021w2v}. Such models do not require annotations and learn to extract contextual representations based on data patterns in the speech signals~\cite{zhang2022bigssl}. State-of-the-art results of various speech recognition tasks can be achieved by finetuning these models or using them as feature extractors. 

With the increasing use of voice-based educational technology, better child ASR systems are needed because speech is one of the mechanisms young children use to interact with devices due to their limited reading and writing abilities. However, child ASR is difficult due to, in part, the lack of large child speech databases. To address this issue, researchers have developed a variety of data augmentation methods by perturbation~\cite{ko2015audio,yeung2021fundamental,yeung2021fundamental,Kat20pitch} or voice conversion~\cite{le2024voicebox,shahnawazuddin20_interspeech}. Another direction is to adopt the pretraining finetuning paradigm, which utilizes the un-annotated data with self-supervised learning~\cite{mohamed2022self,fan2022towards} or the annotated adult data with transfer learning~\cite{shivakumar2020transfer} in the pretraining stage. The knowledge learned in the pretrained models can greatly improve the performance for child ASR. With the recent advances of SFMs, several studies have compared the performance of widely used SFMs on child speech~\cite{jain23_interspeech,attia2023kid,lu2022improving}. However, these studies provide only full finetuning experiments on SFMs. In addition, the speech corpora used in these studies are partitioned differently in each one, making direct performance comparisons difficult.

In this paper, we initiate and present a comprehensive benchmark on the OGI~\cite{shobaki2000ogi} read and MyST spontaneous~\cite{ward2011my} child speech corpora, studying the performance of various SFMs. More importantly, we investigate finetuning strategies for child speech by comparing various data augmentation and parameter-efficient finetuning (PEFT) methods, which are not discussed in previous works. We observe many interesting behaviors of finetuning different speech foundation models. For example, adapter finetuning~\cite{FanA22Draft} is better than full finetuning for large models but vice versa for small models. Our benchmarking study may offer guidance in selecting appropriate models, data augmentation and PEFT strategies to develop robust and accurate child ASR systems. We hope the standard evaluation can lead to fairer comparisons for child ASR research. We also welcome new evaluation sets and new algorithms to include in the benchmark.

\section{Methods in the Benchmark}
In this section, we briefly introduce the methods that are compared in the benchmark, including SFMs, data augmentation and parameter-efficient finetuning (PEFT) techniques.

\subsection{Speech Foundation Models}

\begin{table*}[thp]
  \caption{Details of the speech foundation models used in the benchmark.}
  \label{tab:models}
  \centering
  \begin{tabular}{lccccc}
    \toprule
    \textbf{Model} & Model Architecture & Input Features & Model Size & Sup./Self-sup. & Training Data (hours) \\
    \midrule
    Whisper-\{tiny-large\}     & Encoder-Decoder & Fbank & 39M-1550M & Supervised & 680K \\
    Whisper-largeV3~\cite{Rad23whisper}     & Encoder-Decoder & Fbank & 1550M & Supervised & 1M \\
    Canary~\cite{rekesh2023fast}         & Encoder-Decoder & Fbank  & 1B & Supervised & 85K  \\
    Parakeet-TDT~\cite{rekesh2023fast}   & Transducer &  Fbank & 1.1B & Supervised & 64K \\
    Wav2vec2-Large~\cite{baevski2020wav2vec} & Encoder  & Waveform  & 311M & Self-supervised & 60k  \\           
    HuBERT-Large~\cite{hsu2021hubert}   & Encoder & Waveform & 311M & Self-supervised & 60k\\
    WavLM-Large~\cite{chen2022wavlm}    & Encoder & Waveform & 311M & Self-supervised & 94k \\
    \bottomrule
  \end{tabular}
\end{table*}

Large models trained with large amounts of data have shown great potential to improve the performance of speech recognition tasks. There are two types of SFMs that are: 1) trained with supervised speech-text pairs, such as  Whisper~\cite{Rad23whisper} and Parakeet~\cite{rekesh2023fast}, and 2) trained with unannotated speech data using self-supervised learning, e.g. Wav2vec2.0 \cite{baevski2020wav2vec}, HuBERT \cite{hsu2021hubert}, and WavLM~\cite{chen2022wavlm}. For supervised SFMs, the zero-shot ability of these models is compared as they can directly perform speech recognition tasks. We then conduct in-depth finetuning experiments on the Whisper series (tiny, base, small, medium, large and largeV3). For self-supervised SFMs, finetuning experiments are conducted. The models used in the benchmark are listed in Table~\ref{tab:models} along with details including model architecture, input features, model size and training data. The open-sourced models can be accessed in the OpenASR leaderboard\footnote{\url{https://huggingface.co/spaces/hf-audio/open_asr_leaderboard}}. 

\subsection{Data Augmentation}

Data augmentation methods are commonly used in child ASR systems for alleviating the data scarcity problem, but no systematic comparison between them has been conducted before with all SFM models. Based on the Whisper-small model, we compare several widely-used methods including pitch perturbation (PP)~\cite{patel2011prosodic}, speed perturbation (SP)~\cite{ko2015audio}, vocal tract length perturbation (VTLP)~\cite{jaitly2013vocal}, and SpecAugment (SA)~\cite{park2019specaugment}. Two augmented utterances were created for each utterance as has commonly been done in the literature.

\begin{itemize}
    \item \textbf{Pitch perturbation (PP)} involves altering the fundamental frequency of speech signals while preserving other temporal /spectral features. The pitch is shifted to $n/12$ octave higher or lower for each utterance, where $n$ is randomly sampled from 1 to 12 twice.
    \item \textbf{Speed perturbation (SP)} modifies the speed of speech signals. Two copies of each utterance are created with the perturbation rate of 0.9 and 1.1.
    \item \textbf{Vocal tract length perturbation (VTLP)} involves simulating the effects of variations in vocal tract length by applying frequency warping. The perturbation rate used (0.9 and 1.1) is the same as that used in speed perturbation. 
    \item \textbf{SpecAugment (SA)} randomly masks consecutive frequency bands and time frames, which effectively increases the robustness of the model to time-frequency variations. We use the default SpecAug settings in the Whisper model.
\end{itemize}
We look forward to incorporating new augmentation methods in the benchmark in the future.

\subsection{Parameter Efficient Finetuning (PEFT)}
Parameter-efficient finetuning techniques have become increasingly important when large foundation models are used as model initialization for various tasks~\cite{liu2023sparsely}. These techniques aim to adapt pretrained models to new tasks or domains while minimizing the computational resources required for training. We compare four widely-used PEFT techniques, which are Low Rank Adaptation (LoRA)~\cite{hu2021lora}, adapter tuning~\cite{houlsby2019parameter,FanA22Draft}, prompt tuning~\cite{liu2022p}, and prefix tuning~\cite{li2021prefix}.
\begin{itemize}
    \item LoRA leverages the observation that the matrices of model layers often exhibit low-rank structures. By decomposing weight matrices into low-rank factors and updating only the low-rank factors during fine-tuning, LoRA reduces computational overhead while preserving model performance. We apply LoRA weights to both the query and value-related parameters in each attention layer, with a rank of eight~\cite{hu2021lora}.
    \item Adapter tuning introduces lightweight adapter modules, which are small neural network components inserted between layers of the foundation models. By finetuning only the parameters within the adapters, efficient adaptation is achieved. We used the residual adapters with the bottleneck dimension of 32, which is similar to~\cite{houlsby2019parameter}. The residual adapters are inserted after each block in both the encoder and decoder.
    \item Prompt Tuning prepends randomly initialized prompt vectors to the input sequence and the prompts are optimized through gradient-based methods, allowing the model to directly learn task-specific input representations during fine-tuning. 100 and 20 prompts are inserted in the encoder and decoder inputs, respectively, in our experiments.
    \item Prefix tuning is similar to prompt tuning but prepends the prompts at each layer instead of at the input, bringing more flexibility during finetuning. In our experiments, 50 and 10 prompts are inserted at the beginning of each layer input to both the encoder and decoder modules, respectively.
\end{itemize}
The number of prompts used in prompt tuning and prefix tuning are chosen empirically. By comparing various PEFT methods to full finetuning, we will discover the best finetuning strategy for child ASR when using speech foundation models.

\section{Experiments}
In this section, we present the speech datasets used, experimental setup and results..

\subsection{Child Speech Datasets}

The experiments are conducted on two child speech databases: My Science Tutor (MyST) spontaneous speech corpus~\cite{ward2011my}, and CSLU OGI scripted read speech corpus~\cite{shobaki2000ogi}.

The MyST corpus consists of around 240 hours of transcribed conversational children’s speech (from grade 3 to grade 5), recorded from virtual tutoring sessions in physics, geography, biology, and other topics. Similar to~\cite{attia2023kid}, we identify and filter low quality audio samples by passing the transcribed dataset through the Whisper-largeV2 model. Utterances with WER larger than 50\% or with less than 3 words are removed, resulting in a 133 hours training set. When evaluating the Whisper model, we find that the results are unstable for the test samples that are longer than 30s (the maximum length for training Whisper). Hence, utterances longer than 30s are also removed in both the training and test sets. As a result, the original data splits in MyST corpus are as follows: train (133h), dev (21h), and test (25h).

The CSLU OGI Kids corpus contains around 50 hours of speech by 1100 children (from kindergarten to grade 10) reading from a list that contains either simple words, sentences, or digit strings. The utterances are randomly split into train (70\%), development (15\%) and test (15\%) sets without speaker overlap, which is the same as~\cite{FanA22Draft,fan2021bi}.

The utterance ID list for the two corpora are released as standard evaluations with the training code. 

\subsection{Finetuning and Evaluation Setup}
When finetuning the supervised SFMs, we use the same vocabulary and objective function as those used in the pretraining stage. When finetuning self-supervised SFMs, we use all characters in the transcriptions to create a vocabulary and apply a CTC loss to perform ASR. All results are reported by greedy search decoding without any external language model. The PEFT and DA experiments are not investigated for the OGI corpus because of the page limitation.

\subsection{Zero-shot Performance for the Supervised SFMs}
\label{exp:zero-shot}

\begin{table}[tp!]
  \caption{Zero-shot performance of the supervised speech foundation models in terms of WER. Bold numbers are the best performance among the supervised SFMs.}
  \label{tab:zero-shot}
  \centering
  \begin{tabular}{lccccc}
    \toprule
    \multirow{2}{*}{Model} & \multirow{2}{*}{\shortstack{Model\\ Size}} & \multicolumn{2}{c}{MyST} &  \multicolumn{2}{c}{OGI} \\
    \cline{3-4} \cline{5-6}  
     &  & dev & test & dev & test \\
    
    \midrule
    Whisper-tiny      & 39M  & 18.5 & 20.6 & 40.1 & 53.8 \\
    Whisper-base      & 74M  & 15.6 & 16.8 & 36.8 & 38.0 \\
    Whisper-small     & 242M & 14.4 & 13.4 & 21.2 & 25.4 \\
    Whisper-medium    & 769M & 13.3 & 13.1 & 18.8 & 20.8 \\
    Whisper-large     & 1.55B & 14.4 & 12.5 & 21.2 & 22.9 \\
    Whisper-largeV3   & 1.55B & 12.3 & 12.6	& 14.9 & 19.9\\
    Canary            & 1.0B & \textbf{9.3} & \textbf{9.5} & 14.8 & 18.2 \\
    Parakeet-rnnt       & 1.1B & 10.7 & 11.1 & \textbf{14.3} & \textbf{16.7} \\
    \bottomrule
  \end{tabular}
\end{table}

Since supervised speech foundation models are trained with ASR loss, we first compare their zero-shot abilities on child speech. Top performing models from the OpenASR benchmark for adult speech are selected for comparisons. The results are presented in Table~\ref{tab:zero-shot}. The Canary and Parakeet models have been shown to perform better than Whisper, on average, on the adult speech benchmark~\cite{open-asr-leaderboard}. The same conclusions can be drawn here for child speech, which is surprising because the Whisper models are trained with more data than Canary and Parakeet (training data sizes are shown in Table~\ref{tab:models}). Considering that many of the data for Whisper training is weakly-supervised, we conclude that data quality is sometimes more important than the size of data for obtaining a robust supervised speech foundation model, which has also been observed for large language models~\cite{zhou2024lima}.

\subsection{Foundation Models with Finetuning}
\label{exp:sup_and_ssl}

\begin{table}[tp!]
  \caption{WER comparisons of finetuning supervised and self-supervised speech foundation models. Note finetuning on each corpus separately. Supervised and self-supervised SFMs are finetuned with 2GPUs for 4k and 12k steps, respectively.}
  \label{tab:sup_and_ssl}
  \centering
  \begin{tabular}{lccccc}
    \toprule
    \multirow{2}{*}{Model} &  \multicolumn{3}{c}{MyST} &  \multicolumn{2}{c}{OGI}\\
    \cline{2-4} \cline{5-6}
    ~ & Dev & Test & Time & Dev & Test \\
    \midrule
    \multicolumn{5}{c}{Supervised SFM} \\
    \hline
    Whisper-tiny    & 11.6 & 11.6 & 2.0h & 2.7	& 3.0 \\
    Whisper-base    & 9.1  & 10.4 & 2.5h & 2.0 & 2.3 \\
    Whisper-small   & 8.4  & 9.3  & 6.0h & 5.0 & 1.8 \\
    Whisper-medium  & 8.4  & \textbf{8.9}  & 8.0h & \textbf{1.6}   & 1.5	  \\
    Whisper-large   & \textbf{8.2} & 13.0 & 9.2h  & 1.8 & 1.7   \\
    Whisper-largeV3 & 8.5  & 9.1 & 13.0h & \textbf{1.6}	& \textbf{1.4}  \\
    \hline
    \multicolumn{5}{c}{Self-supervised SFM} \\
    \hline
    Wav2vec2.0      & 10.6 & 11.1 & 10.5h & 2.1 & 2.5 \\
    HuBERT          & 10.5 & 11.3 & 10.5h & 2.2	& 2.5 \\
    WavLM           & 9.6  & 10.4 & 13.5h & 1.7	& 1.8 \\
    \bottomrule
  \end{tabular}
\end{table}

In addition to the supervised foundation models, self-supervised foundation models are also widely used for child ASR. We compare the full-finetuning performance between the two types of foundation models and present the results in Table~\ref{tab:sup_and_ssl}. The results show that supervised SFMs can achieve better performance than the self-supervised SFMs after finetuning with similar model parameters (e.g. Whisper-small with 242M and WavLM with 311M parameters). Among the most widely used SSL models, WavLM achieves the best performance because it used more data and included a masked reconstruction from noisy and multi-talker speech data during pretraining. Note that an advantage of the SSL models are that they might also be robust to other speech tasks because the SSL loss is not specifically designed for ASR. We don't compare this ability of SSL models since we are mainly focusing on the ASR task. The full finetuning results for the Canary and Parakeet models are as follows: Canary (MyST dev: 8.6, MyST test: 9.2, OGI dev: \textbf{1.4}, OGI test: 1.5); Parakeet (MyST dev: \textbf{7.9}, MyST test: \textbf{8.5}, OGI dev: 1.8, OGI test: 1.8). Due to the recent release of these models, PEFT and DA experiments were not explored in detail.

\subsection{Comparisons of Data Augmentation Methods}
\label{exp:data_aug}

\begin{table}[thp]
  \caption{WER comparisons of different data augmentation methods on MyST dataset using the Whisper-small model. PIF stands for perturbation invariant training. $\text{x3}$ indicates three copies of the original data for augmentation.}
  \label{tab:dataaug}
  \centering
  \resizebox{0.85\columnwidth}{!}{%
  \begin{tabular}{lccc}
    \toprule
    Whisper-small & Augmentation & MyST-dev &  MyST-test \\
    
    \midrule
    Baseline    & no     & 14.4 & 13.4  \\
    \hline
    \multirow{8}{*}{Finetuning}  & no     & 8.4 & 9.3 \\
                & PP (x3) & 8.6 & \textbf{8.8} \\
                & VTLP (x3)   & 8.6 & 9.0 \\
                & SP (x3) & \textbf{8.1} & 8.9 \\
                & SA & 8.2 & 9.0 \\
                & SA + PP & 8.2 & 8.9\\
                & SA + VTLP & \textbf{8.1} & 9.0 \\
                & SA + SP & 8.3 & 8.9 \\
    \hline
    \multirow{2}{*}{PIF}  & VTLP (x3) & 8.3 & 9.0 \\
                & PP (x3) & 8.3 & 8.9 \\
    \bottomrule
  \end{tabular}%
}
\end{table}

Data augmentation (DA) is an important technique to deal with low-resource tasks, such as child ASR. However, previous works either used private data or conducted experiments with their own settings, making the comparisons of different methods difficult. In addition, previous DA methods are proposed based on training from scratch. It is unknown whether these methods improve the performance when using SFMs. To address this issue, we made a comparison of different DA methods and explored their role in finetuning SFMs. The experimental results on MyST dataset using Whisper-small model are shown in Table~\ref{tab:dataaug}. The reason we use the Whisper-small model is that it is computationally efficient given our limited number of GPUs, and achieves a reasonable WER on child speech. We can observe from the table that different augmentation methods achieve similar WER improvements compared to the finetuning baseline. Interestingly, the combination of two DA methods does not provide further gains compared to using only one method. This is slightly different from the conclusion in~\cite{yeung2021fundamentalconf} when the model is trained from scratch. This might be because the SFM itself is already robust to some variations created by the DA methods. Note that F0-based data augmentation in~\cite{yeung2021fundamentalconf} achieves similar performance to pitch perturbation. 

The improvements of PP and VTLP are not stable, and we propose a perturbation invariant finetuning (PIF) technique to stabilize the VTLP and PP. Specifically, an additional distance loss between the encoder outputs of original and perturbed utterance is added as a regularization for finetuning. The results in Table~\ref{tab:dataaug} show that PIF can lead to more consistent improvements of perturbation methods on the MyST-dev and MyST-test sets. PIF is only used for VTLP and PP as they are not stable when FT on kids speech while other DAs are stable.

\subsection{Comparisons of Parameter Efficient Finetuning}
\label{exp:peft}

\begin{table}[tp!]
  \caption{WER comparisons of different parameter efficient finetuning (PEFT) methods on MyST dataset using the Whisper-small model. Params indicates the number of updated parameters during finetuning. Enc. and Dec. represents finetuning encoder and decoder only, respectively.}
  \label{tab:peft}
  \centering
  \resizebox{0.85\columnwidth}{!}{%
  \begin{tabular}{lcccc}
    \toprule
    Model & PEFT & MyST-dev &  MyST-test & Params\\
    \midrule
    Baseline    & no     & 14.4 & 13.4 & 0 \\
    Full-FT     & no     & \textbf{8.4} & 9.3 & 242M\\
    \hline
    \multirow{6}{*}{\shortstack{Whisper\\-small}}
                & Enc. & 9.0	& \textbf{9.2} & 88M \\
                & Dec.   & 8.9 & 9.5 & 154M \\
                & Prompt~\cite{li2023prompting} & 10.4	& 10.4 & 92k \\
                & Prefix~\cite{li2021prefix} & 8.9 & 10.2 & 541k \\
                & LoRA~\cite{hu2021lora} & 9.1 & 9.6 & 917k \\
                & Adapter~\cite{FanA22Draft} & \textbf{8.4} & 9.3 & 1.29M \\
    \bottomrule
  \end{tabular}%
}
\end{table}

When speech foundation models are large, full finetuning with the entire model parameters would be difficult because of the high GPU memory costs. Parameter efficient finetuning (PEFT) can retain the performance of full tuning but update less parameters during the finetuning stage. We compare several widely used PEFT methods in the NLP area on Whisper-small model on the MyST dataset and present the results in Table~\ref{tab:peft}. It can be seen from the table that adapter tuning achieves similar performance compared to the full finetuning while having only 1.29M parameters for updates. Note that the initialization of the adapters are important for good performance of adapter tuning. For example, the inserted adapter module should be equivalent to the identity function at the start of the finetuning. However, LoRA, the most popular PEFT method in the area of NLP, achieves worse performance than the full finetuning. Prompt and prefix tuning behave not as good as LoRA and adapter FT might be because they alter the positional information of the speech sequence and restrict the model capacity for learning from finetuning data. 


\subsection{Impact of Model size on PEFT performance}

\begin{figure}[tp]
    \centering
    \subfloat[MyST-test]{
    \includegraphics[height=0.20\textwidth,width=0.24\textwidth]{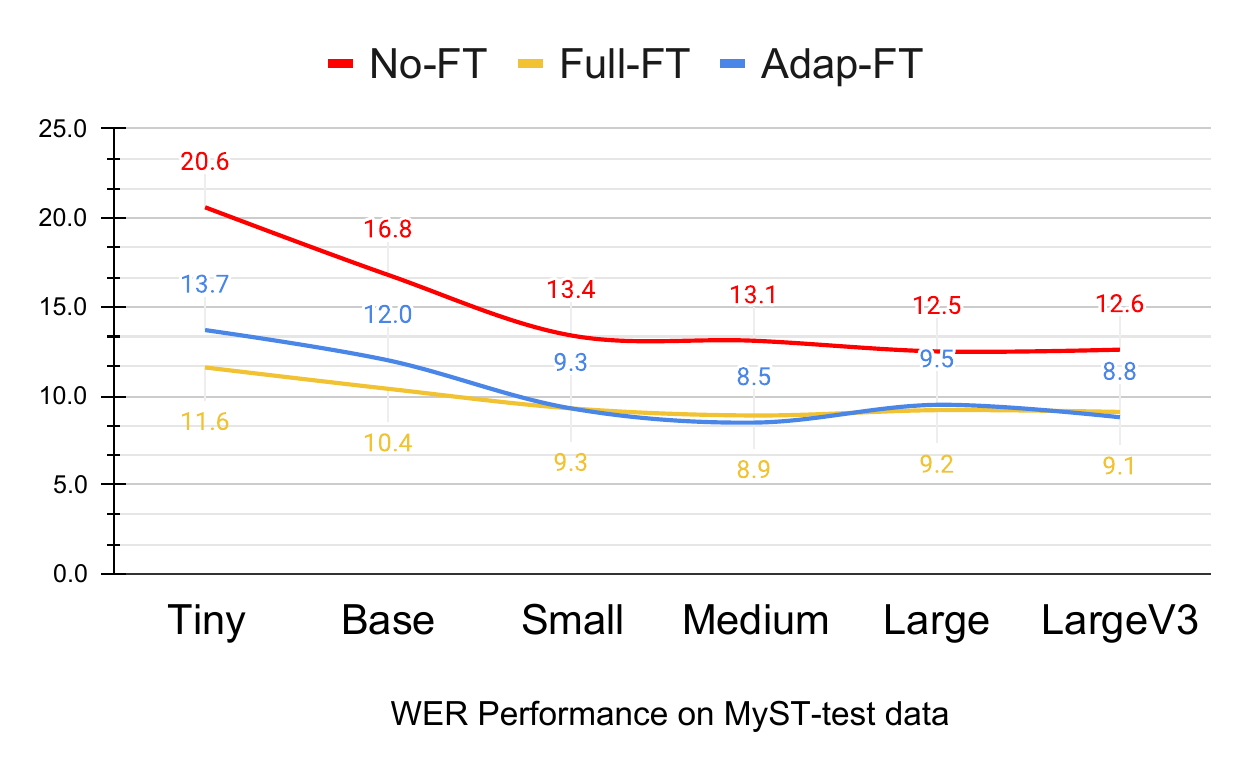}} %
    \subfloat[OGI-test]{
    \includegraphics[height=0.20\textwidth,width=0.24\textwidth]{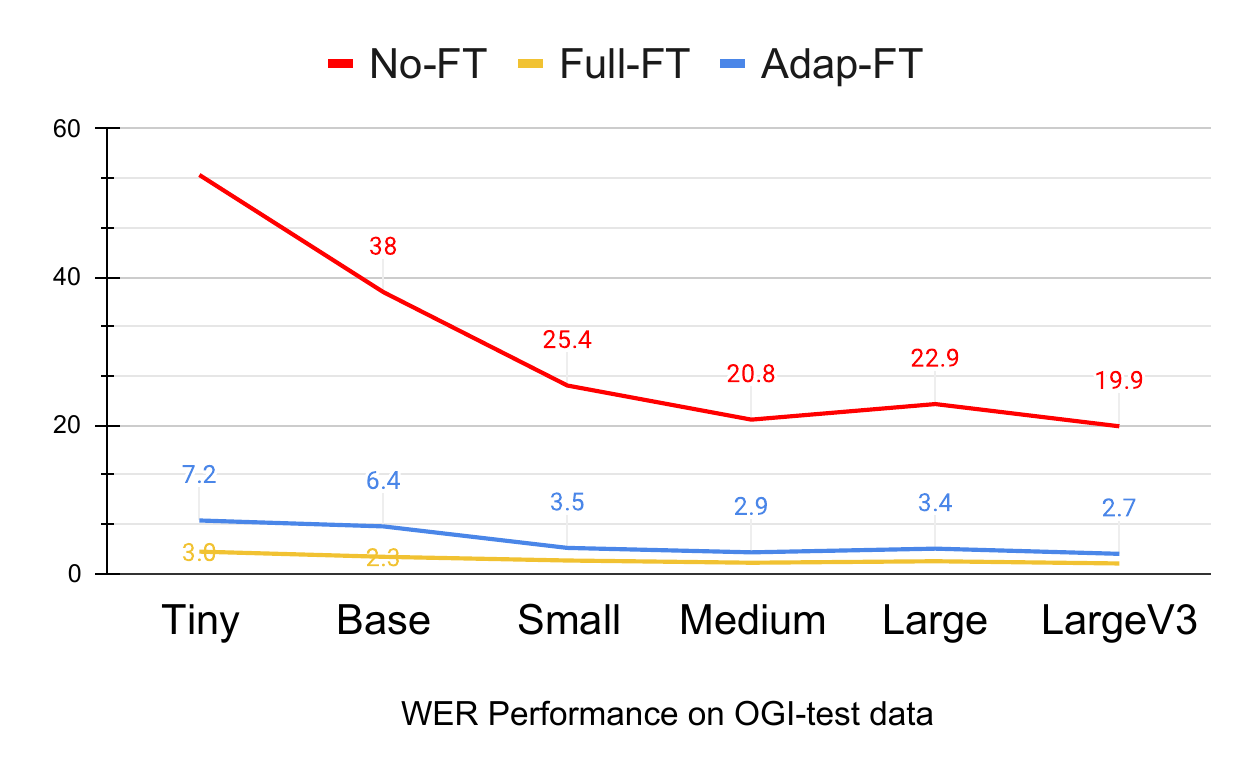}}%
    
    \caption{The impact of the Whisper model size for full and adapter finetuning (Adap-FT in the figure) on WER. The model size of each whisper model can be found in Table~\ref{tab:models}. } %
    \label{fig:size_impact}%
\end{figure}

As shown in Table~\ref{tab:sup_and_ssl}, the WER of the Whisper model decreases when the model size increases. We further explore whether the model size would affect the performance of PEFT, specifically adapter tuning because it behaves better than other PEFTs as shown in Table~\ref{tab:peft}. The results of both full finetuning and adapter finetuning on MyST and OGI test data are plotted in Figure~\ref{fig:size_impact}. We can observe from the figure that the adapter tuning does not work as well as the full finetuning for small models. However, when the model size increases, the gap between adapter tuning and full finetuning decreases. For example, the adapter tuning matches the performance of the full finetuning for the Whisper-largeV3 on the MyST-test data. This interesting behavior provides us with guidance on how to select the appropriate finetuning strategy. That is, performing full finetuning for small models and PEFT for large models. It would also be interesting to investigate the impact of the model size for data augmentation methods, which will be included in future work.

\section{Conclusions and Future Work}

In this paper, we presented the first benchmark for child ASR with a comparison of various speech foundation models, such as Whisper, Canary, Parakeet, Wav2vec2.0, HuBERT, and WavLM. We found that the Canary and Parakeet models are better than Whisper models on child speech with much less training data, indicating the data quality is sometimes more important than the data quantity. As expected, supervised SFMs performed better than the self-supervised SFMs after finetuning. Moreover, we investigated finetuning strategies by comparing various data augmentation (pitch perturbation, speed perturbation, VTLP and SpecAugment) and parameter-efficient finetuning (PEFT) methods (prompt tuning, prefix tuning, adapter tuning, and LoRA). To stabilize the finetuning using the augmented data, we propose a perturbation invariant finetuning (PIF) loss as a regularization. Various parameter-efficient finetuning (PEFT) strategies were compared, and we observed that the behaviors of PEFT are different when the model size increases. For example, PEFT performed better than full finetuning for large models but worse for small models. This study may offer guidance in selecting appropriate models, data augmentation and PEFT strategies to develop robust child ASR systems. 

Future work will include: 1) Evaluations on other child speech datasets; 2) Comparisons with new data augmentation methods; 3) Evaluations of other open-sourced speech foundation models, such as SeamlessM4T~\cite{barrault2023seamless}, OWSM~\cite{peng2023reproducing} and W2VBERT2.0~\cite{chung2021w2v}; 4) Migration of models not supported in Huggingface, e.g. the Canary and Parakeet models developed using the NeMo~\cite{kuchaiev2019nemo} framework, since our finetuning code is implemented based on Huggingface.

\footnotesize
\bibliographystyle{IEEEtran}
\bibliography{mybib}

\end{document}